\documentclass{iaus}
\usepackage{graphicx}

\title[Central stars of planetary nebulae: The white dwarf connection] %% give here short title %%
{Central stars of planetary nebulae: \\The white dwarf connection}

\author[Klaus Werner]   %% give here short author list %%
{Klaus Werner}

\affiliation{Institute for Astronomy and Astrophysics, Kepler Center for Astro
  and Particle Physics, Eberhard Karls University, Sand 1, 72076 T\"ubingen, Germany \\ email: {\tt werner@astro.uni-tuebingen.de} }

\pubyear{2012}
\volume{283}  %% insert here IAU Symposium No.
\pagerange{1--8}
\jname{Planetary Nebulae: An Eye to the Future}
\editors{A. Manchado, L. Stanghellini, \& D. Sch\"onberner, eds.}

\def\kpd{KPD\,0005+5106}
\def\hh{H\,1504$+$65}
\renewcommand{\etal}{et\,al.\ }
\newcommand{\logg}{\mbox{$\log g$}}
\newcommand{\Teff}{\mbox{$T_\mathrm{eff}$}}
\newcommand{\lppr}{\stackrel{<}{\scriptstyle \sim}}
\newcommand{\lappr}{\raisebox{-0.4ex}{$\lppr $}}

\newcommand{\ion}[2]{\mbox{#1\,{\sc #2}}}

\begin{document}

\maketitle

\begin{abstract}
This paper is focused on the transition phase between central stars and white
dwarfs, i.e. objects in the effective temperature range 100\,000 --
200\,000~K. We confine our review to hydrogen-deficient stars because the common
H-rich objects are subject of the paper by Ziegler \etal in these
proceedings. We address the claimed iron-deficiency in PG1159 stars and [WC]
central stars. The discovery of new \ion{Ne}{vii} and \ion{Ne}{viii} lines in
PG1159 stars suggests that the identification of \ion{O}{vii} and \ion{O}{viii}
lines that are used for spectral classification of [WCE] stars is wrong. We then
present evidence for two distinct post-AGB evolutionary sequences for
H-deficient stars based on abundance analyses of the He-dominated O(He) stars
and the hot DO white dwarf \kpd. Finally, we report on evidence for an
H-deficient post-super AGB evolution sequence represented by the hottest known,
carbon/oxygen-atmosphere white dwarf \hh\ and the recently discovered
carbon-atmosphere ``hot DQ'' white dwarfs.  \keywords{stars: abundances, stars:
AGB and post-AGB, stars: atmospheres, stars: early-type, stars: evolution, white
dwarfs, planetary nebulae: general, ultraviolet: stars}
%% add here a maximum of 10 keywords, to be taken form the file <Keywords.txt>
\end{abstract}

\firstsection % if your document starts with a section,
              % remove some space above using this command.
\section{Introduction}

\cite[M\'endez \etal (1986)]{mendez:86} have introduced the O(He) and O(C)
designations (besides others) to classify the optical spectra of hot
hydrogen-deficient central stars. The O(He) stars have almost pure \ion{He}{ii}
absorption line spectra whereas the O(C) stars additionally show strong lines
from \ion{C}{iv} and sometimes \ion{O}{vi}. While the  O(He) designation is
still in use, the O(C) stars are today more commonly called PG1159 stars after
their prototype PG1159$-$035. They also comprise some of the stars that were
previously termed ``\ion{O}{vi}'' central stars. PG1159 stars are thought to be
the progeny of the Wolf-Rayet type central stars (spectral type [WC]), while future
evolution will turn them into non-DA white dwarfs. In fact, the picture is a bit
more complicated, because a few PG1159 stars show traces of H and these should
become DA white dwarfs. The existence of remnant hydrogen can be explained by a
special variant of the late He-shell flash (or born-again star) scenario, which
is invoked to be the origin of the H-deficient chemistry.

PG1159 and [WC] stars are particularly useful to test AGB star nucleosynthesis
models, because the late He-shell flash that the stars have suffered laid bare
the intershell matter between the H and He burning shells. The observed surface
abundances can be directly compared to model predictions for the chemical
composition of the AGB star intershell region, where nucleosynthesis of heavy
elements proceeds. In this context, one particular problem is the unexpectedly
strong iron deficiency claimed for PG1159 and [WC] stars. This is addressed in
Sect.\,\ref{sect_fe}. We will also highlight some results on abundance
determinations of other metals in PG1159 stars
(Sect.\,\ref{sect_metals}). During the course of neon line identifications and
abundance determinations it was discovered that the identification of
ultrahigh-ionization spectral lines of oxygen -- that are commonly used to
classify early-type [WC] stars -- is wrong (Sect.\,\ref{sect_ne}).

While we think that the O(C) (thus PG1159) stars are part of the sequence [WC]
$\longrightarrow$ PG1159 $\longrightarrow$ non-DA WD, the evolutionary context
of O(He) stars is less clear. They are not explained by the late He-shell flash
scenario and there is additional evidence that they indeed belong to a distinct
H-deficient post-AGB sequence (see Sect.\,\ref{sect_ohe}).

Finally, in Sect.\,\ref{sect_h15} we dwell on rather exotic white dwarfs with
atmospheres devoid of H and He, which might be the progeny of super-post AGB
stars, i.e. red giants that burned carbon.

\begin{figure}[t]
%\begin{center}
 \includegraphics[width=\textwidth]{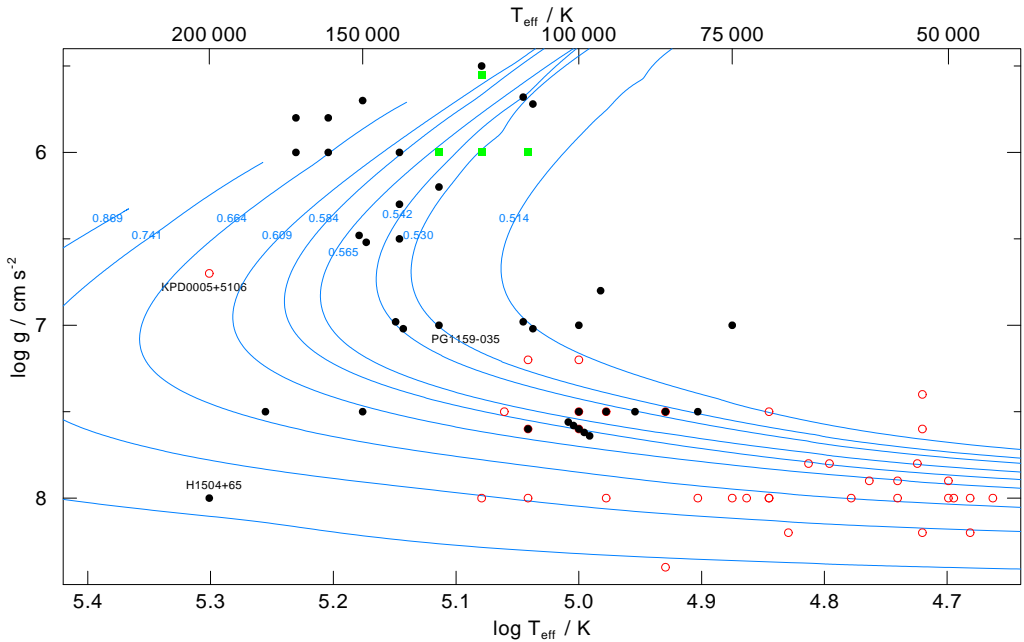}
%\end{center}
 \caption{Location of PG1159 stars (filled circles), DO white dwarfs (open
   circles), and O(He) stars (squares) in the g-\Teff\ diagram. Atmospheric
   parameters for PG1159s and DOs were taken from compilations in \cite[Werner
   \& Herwig (2006)]{werner:06} and \cite[Althaus \etal (2009)]{alt:09},
   respectively, with some improvements according to more recent work. O(He)
   parameters are from Reindl \etal (these proceedings). Evolutionary tracks for
   non-DA white dwarfs are from \cite[Althaus \etal (2009)]{alt:09} and they are
   labeled with the WD mass in M$_\odot$.  }
   \label{fig:gteff}
\end{figure}

\section{PG1159 and [WC] stars}

Recent comprehensive reviews on stellar and atmospheric parameters of [WC] and
PG1159 stars can be found in \cite[Crowther (2008)]{crowther:08} and
\cite[Werner \etal (2008)]{werner:08}, respectively. The stellar evolution
context was summarized by \cite[Werner \& Herwig (2006)]{werner:06}.

In the HRD, [WC] stars are evolving along the constant-luminosity, horizontal
part of post-AGB tracks towards high effective temperature. They cover the range
of \Teff\ $\approx$ 20\,000\,K--150\,000\,K. Along the way, the stars shrink
such that their surface gravity increases from roughly \logg\ = 3 to 6. With
decreasing mass-loss rate, the stars turn into PG1159 stars. The latter populate
the region where the stars evolve around the ``knee'' of the tracks, reaching
the maximum effective temperature and subsequently cooling along the WD
sequence. Their gravity increases from \logg\ = 5.5 to 8.  The hottest PG1159
stars have \Teff\ near 200\,000\,K, while the most evolved ones are observed
at about \Teff\ = 75\,000\,K. At this region, the stars turn into He-rich white
dwarfs (or H-rich in the case some trace H was left) because of gravitational
settling of heavy elements. Figure~\,\ref{fig:gteff} shows the position of all
analyzed PG1159 stars, O(He) stars, and hot non-DA white dwarfs (spectral type
DO) in the g--\Teff\ diagram.  Note that roughly every other PG1159 star has no
associated PN, probably because of its advanced evolutionary state.

The most abundant elements in [WC] and PG1159 atmospheres are He, C, and
O. Their relative abundance varies strongly from star to star. For example, a
typical mass ratio is displayed by the PG1159 prototype: He/C/O = 33/48/17.

As already mentioned in the introduction, abundances of other metals are
interesting to compare with nucleosynthesis models. In the next subsections, we
concentrate on some specific highlights and problems, focusing on PG1159 stars.

\begin{figure}[t]
\begin{center}
 \includegraphics[width=0.9\textwidth]{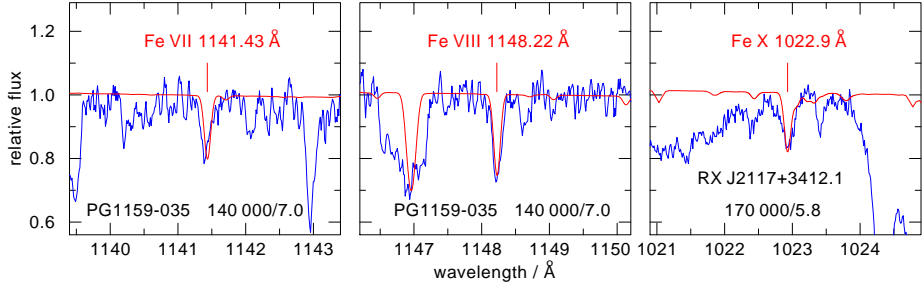}
 \caption{High-ionization iron lines in PG1159 stars. \emph{Left two panels:} \ion{Fe}{vii}
and \ion{Fe}{viii} lines in the PG1159 prototype. \emph{Right:} \ion{Fe}{x} line
    in the central star RX\,J2117+3412.1. Numbers in panels denote \Teff\ and
    \logg\ of the overplotted model profiles. The Fe abundance in the models is
    solar.  }
   \label{fig:fe}
\end{center}
\end{figure}

\subsection{Are PG1159 and [WC] stars iron deficient?}\label{sect_fe}

Based on the non-detection of iron lines in any PG1159 star, an iron deficiency
up to about one dex was claimed for a number of objects (e.g. \cite[Jahn \etal
2007]{jahn:07}). This is unexplained by stellar models, because they predict an
only marginal Fe reduction by 10\% due to neutron captures. An unexpectedly
strong Fe depletion was also reported for [WC] stars (e.g. \cite[Crowther \etal
1998]{crowther:98}).

For the PG1159 stars this problem could be resolved recently by the discovery of
\ion{Fe}{viii} and \ion{Fe}{x} lines \cite[(Werner \etal 2010,
2011; Fig.\,\ref{fig:fe})]{werner:10,werner:11} in eight of the hottest objects. The derived iron
abundances are solar, being in line with stellar models within the analysis
error limits. The solar iron abundance in one particular object (PG\,1424+535)
is consistent with its solar argon abundance \cite[(Werner \etal 2007)]{werner:07}. Argon is an
independent metalicity indicator because its abundance remains unchanged in AGB
star nucleosynthesis (see, e.g, Lugaro in these proceedings).

In the case of [WC] stars the problem remains. In a first attempt
to model \ion{Fe}{viii} and \ion{Fe}{x} lines in a [WC], \cite[Keller \etal (2011 and
these proceedings)]{keller:11} arrive at a significant iron deficiency (down to
0.3 solar) for the [WCE] NGC\,6905. All the more astonishing is their concurrent result of a
ten times solar argon abundance.

We note that the newly discovered \ion{Fe}{viii} lines are also present in many
H-rich central stars and DO white dwarfs \cite[(Werner \etal 2011)]{werner:11} and together
with \ion{Fe}{vii} lines they serve as a new sensitive temperature diagnostics
(Ziegler \etal and Mahsereci \etal in these proceedings).

\begin{figure}[t]
\begin{center}
\includegraphics[width=0.9\textwidth]{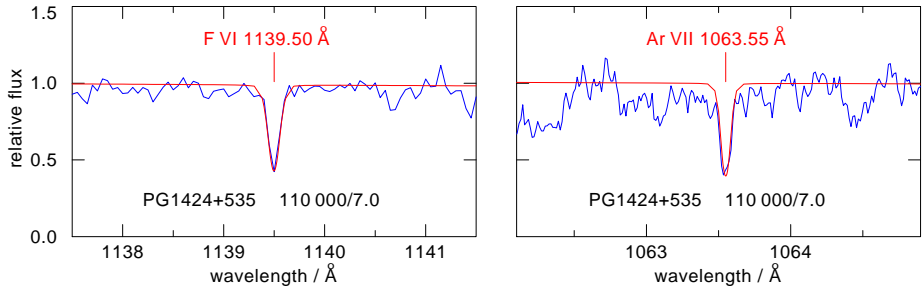}
\end{center}
\caption{Fluorine line \emph{(left panel)} and argon line \emph{(right panel)} in the PG1159 star PG1424+535. The fluorine
  and argon  abundances in the overplotted model are 200 times solar and solar,
  respectively.}
   \label{fig:pg1424}
\end{figure}                                                                                        

\subsection{Other trace metals in PG1159 stars: N, F, Ne, Si, P, S, Ar}\label{sect_metals}

\begin{figure}[t]
\begin{center}
\includegraphics[width=0.9\textwidth]{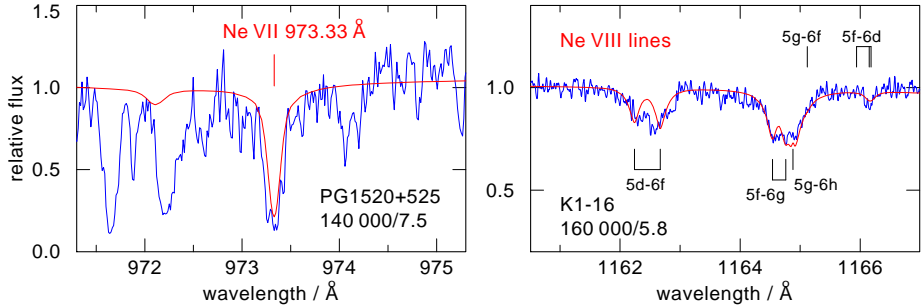}
\end{center}
\caption{Highly ionized neon lines in PG1159-type central stars. \emph{Left:} 
\ion{Ne}{vii}~$\lambda$\,973.3~\AA\ in PG1520+525.
 \emph{Right:}  \ion{Ne}{viii} lines in  K\,1-16. The neon abundance in the
 overplotted models is 2\% by mass.}
   \label{fig:ne}
\end{figure}

Far-UV spectroscopy with \emph{FUSE} augmented by UV spectroscopy with HST was
seminal for first-time detection of particular elements and ionization stages in
PG1159 stars. Besides iron (\ion{Fe}{vii}, \ion{Fe}{viii}, \ion{Fe}{x}), the
elements fluorine (\ion{F}{v}, \ion{F}{vi}; left panel of
Fig.\,\ref{fig:pg1424}), phosphorus (\ion{P}{v}), neon (\ion{Ne}{vi--viii},
Fig.\,\ref{fig:ne}), and argon (\ion{Ar}{vii}; right panel of
Fig.\,\ref{fig:pg1424}) were identified for the first time. From silicon, which
is usually detected through its \ion{Si}{iv} resonance doublet, very high
ionization stages were discovered (\ion{Si}{v--vii}).

Most trace metal abundances in PG1159 stars are consistent with predictions from
nucleosynthesis models. For example, extreme fluorine abundances were determined
in some stars (up to 200 times solar), while argon is solar. After the iron
abundance problem was solved, the largest remaining discrepancy is for
sulfur. In four out of five stars, we found unexpectedly strong depletions down
to about 0.1 times solar or less \cite[(e.g. Jahn \etal 2007)]{jahn:07}. It is
remarkable, that such a sulfur anomaly is also encountered at abundance analyses
of PNe \cite[(Henry \etal 2006)]{henry:06}.

The nitrogen abundance is an important marker for the event leading to
hydrogen-deficiency in PG1159 and [WC] stars. The absence (less than 0.1\%, by
mass) or presence (few \%) of N is a reliable indicator of a late or very late
thermal pulse (i.e. final pulse occurred in H-burning post-AGB state or WD
cooling phase, respectively).

Neon deserves special attention. Its observed abundance is  2\%
(i.e. $\approx$\,15 times solar), in agreement with expectations from stellar
models. During evolution, all CNO is mostly transformed into $^{14}$N, which is
subsequently transformed to $^{22}$Ne by $\alpha$ captures. The discovery of
neon by the identification of \ion{Ne}{vii} and \ion{Ne}{viii} in UV spectra and
in the optical wavelength region led to the conclusion that previous
identifications of ultrahigh-ionization lines in WR stars (Pop.\ I and II) and
in PG1159 stars must be revised.

\subsection{Neon vs. oxygen: wrong identification of [WC] spectral classification lines}\label{sect_ne}

In order to classify the hottest [WR] stars, in particular the [WO] (which are a
continuation of [WC] to higher effective temperatures), line ratios of
\ion{O}{vi}/\ion{O}{vii}/\ion{O}{viii} are used \cite[(Acker \& Neiner
2003)]{acker:03}. Lines from \ion{O}{vii} and \ion{O}{viii} are called
``ultrahigh ionization'' lines \cite[(Barlow \etal 1980)]{barlow:80}, because
temperatures of the order 10$^6$\,K are required to excite the respective energy
levels. These are, however, not attained in the atmospheres unless e.g. shocks
are invoked.

We have shown that the \ion{O}{vii} and \ion{O}{viii} line identifications are
probably wrong (Werner \etal 2007). All features can be ascribed to
\ion{Ne}{vii} and \ion{Ne}{viii} lines, respectively. This identification is
much more natural because the neon lines are excited ordinarily in hot star
atmospheres without invoking shocks, because and neon is an abundant element in
[WC] and PG1159 stars. As a consequence, the putative \ion{O}{vi}/\ion{O}{vii}
line ratios for example  are in fact ratios of \ion{O}{vi} to \ion{Ne}{vii}
lines. The current spectral classification scheme is therefore affected by the
O/Ne abundance in the star and not only by the temperature.

Other features that are commonly ascribed to ultrahigh ionization lines from
carbon (\ion{C}{v} and \ion{C}{vi}) are in fact lines from ordinarily excited
\ion{N}{v} and \ion{O}{vi}, respectively. Table~\ref{tab_lines} summarizes the
new line identifications.

\begin{table}
\begin{center}
\caption{
Old ultrahigh ionization line identifications and revised identifications.
} 
\label{tab2} 
\begin{tabular}{cllll}  
\noalign{\smallskip} \hline  \noalign{\smallskip} Wavelength (\AA)&
\multicolumn{2}{l}{Ultrahigh-ionization identification}   &
\multicolumn{2}{l}{New identification} \\ \hline  \noalign{\smallskip} 1932
& \ion{O}{viii} & $n=7\rightarrow  8$ & \ion{Ne}{viii} & $n=7\rightarrow 8$ \\
2977             & \ion{O}{viii} & $n=6\rightarrow  7$ & \ion{Ne}{viii} &
$n=6\rightarrow 7$ \\ 3893             & \ion{O}{vii}  & $n=7\rightarrow  8$ &
\ion{Ne}{vii}  & 3p\,$^3$P$^{\rm o}\rightarrow$ 3d\,$^3$D\\ &               &
& plus \ion{Ne}{vii} & $n=7\rightarrow  8$ \\ 4340             & \ion{O}{viii} &
$n=8\rightarrow  9$ & \ion{Ne}{viii} & $n=8\rightarrow 9$ \\ 4500             &
\ion{C}{vi}   & $n=8\rightarrow 10$ & \ion{O}{vi}    & $n=8\rightarrow 10$ \\
4555             & \ion{O}{vii}  & $n=9\rightarrow 11$ & \ion{Ne}{vii}  &
$n=9\rightarrow 11$ \\ 4945             & \ion{C}{v}    & $n=6\rightarrow  7$ &
\ion{N}{v}     & $n=6\rightarrow  7$ \\   5290             & \ion{C}{vi}   &
$n=7\rightarrow  8$ & \ion{O}{vi}    & $n=7\rightarrow 8$ \\ 5665             &
\ion{O}{vii}  & $n=8\rightarrow  9$ & \ion{Ne}{vii}  & $n=8\rightarrow 9$  \\
6068             & \ion{O}{viii} & $n=9\rightarrow 10$ & \ion{Ne}{viii} &
$n=9\rightarrow 10$ \\ \noalign{\smallskip} \hline
\end{tabular} \label{tab_lines}
\end{center}
\end{table}

\section{Two distinct post-AGB evolutionary sequences for H-deficient stars?}\label{sect_ohe}

The O(He) stars are a small group of four objects (two have an associated PN)
with parameters close to \Teff\ = 120\,000\,K and \logg\ = 6 \cite[(Rauch \etal
1998, 2008]{ra:98,ra:08}, and Reindl \etal in these proceedings;
Fig.\,\ref{fig:gteff}). They have helium-dominated atmospheres and cannot be
explained by a late flash like the PG1159 and [WC] stars, because evolutionary
models always predict high C abundances. One can therefore argue that they are
representatives of a distinct post-AGB sequence. If so, which stars are possible
progenitors and successors?

It could be that some of the O(He) stars are evolved R CrB stars (\Teff\ around
7000\,K) and we are currently investigating how metal abundance patterns compare
(see Reindl \etal in these proceedings). Even if the relationship is identified,
the question about their origin remains unanswered. R CrB stars may be the
outcome of a double-degenerate merger, i.e. the coalescence of a carbon-oxygen
WD with a helium WD. Merger simulations indicate that the resulting chemical
abundance patterns are in qualitative agreement with observed abundances in R
CrB stars, although it is not clear to what extent the resulting metal
abundances are determined by the intershell composition of the C-O WD progenitor
or by nucleosynthesis during merging \cite[(Jeffery \etal 2011, Longland \etal
2011)]{je:11,lo:11}.  Helium-rich objects that bridge the large \Teff\ gap
between the R CrB and O(He) stars could be the extreme He-B stars (EHeB, \Teff\
around 20\,000\,K) and a handful known low-gravity sdO stars (\Teff\ around
70\,000\,K, see e.g. \cite[Jeffery 2008]{je:08}).

Two of the O(He) stars have a significant amount of hydrogen (H/He = 0.1 and 0.5
by number in HS\,1522+6615 and LoTr\,4, respectively; \cite[Rauch \etal 1998;
Reindl \etal in these proceedings]{ra:98}). Also, a few R CrB stars have much
hydrogen (e.g. H/He = 0.5 in V854 Cen; \cite[Rao \& Lambert 1996]{rao:96}), as
well as some EHeB stars. This is in conflict with the merger scenario because
one would expect very little or no remaining hydrogen. It is speculated that an
alternative origin are post-early AGB stars, i.e. rather low-mass objects that
experience their first thermal pulse after departure from the AGB (\cite[Miller
Bertolami \& Althaus 2006]{mi:06}).

Immediate progenitors of the O(He) stars might be [WN] central stars. Their
existence, however, is still debated (\cite[Todt \etal 2010a]{to:10a}). It is
claimed that the central star of PB\,8 (\Teff\ $\approx$ 50\,000\,K) is indeed a
[WN] (or more precise: a [WN6]/[WC7] type), being He-dominated with a large H
abundance (H/He/C/N/O = 40/55/1.3/2/1.3, mass fractions), and a possible
relation to the O(He) class was discussed (\cite[Todt \etal
2010b]{to:10b}). \cite[Miller Bertolami \etal (2011)]{mi:11} argue that PB\,8
might be the result of a diffusion-induced nova, which is an entirely different
scenario to explain H-deficient post-AGB stars. It could perhaps also apply to
the O(He) stars.

A potential successor of the O(He) stars could be the He-dominated \kpd\ (\Teff\
= 200\,000\,K, \logg\ = 6.7; \cite[Wassermann \etal 2010]{wa:10}). Its metal
abundance pattern reveals strong commonalities with R CrB stars.  Like the
PG1159 stars, with increasing gravity the O(He) stars will evolve into DA or
non-DA WDs depending on whether they retained hydrogen or not.

\section{Evidence for H-deficient post-super AGB evolution}\label{sect_h15}

\begin{figure}[t]
\begin{center}
\includegraphics[width=0.9\textwidth]{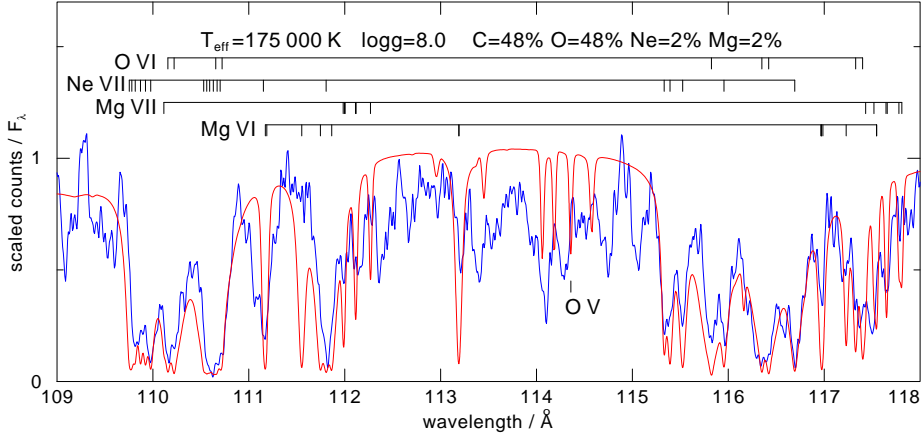}
\end{center}
\caption{Detail from a \emph{Chandra} soft X-ray spectrum of \hh\ and an overplotted
  model with parameters as noted in the panel.}
   \label{fig:h1504}
\end{figure}

\hh\ is the hottest known white dwarf (\Teff\ = 200\,000\,K). Its high surface
gravity (\logg\ = 8) indicates a relatively high mass.
Spectroscopically, it is related to the PG1159 class but it is a distinct
object because it is not only hydrogen-deficient but also helium-deficient. The
atmosphere is primarily composed of carbon and oxygen, by equal amounts
\cite[Werner (1991)]{werner:91}. In addition, a high abundance of neon was
derived \cite[Werner \& Wolff (1999)]{werner:99}. The origin of this exotic
surface chemistry (C = 49\%, O = 49\%, Ne = 2\%, mass fractions) is completely
unclear. We have speculated that \hh\ represents the naked C--O core of a white
dwarf. Another, even more exciting possibility is that we see the eroded C--O
envelope of a O--Ne--Mg white dwarf. This is corroborated by a \emph{Chandra}
soft X-ray spectrum \cite[(Werner \etal 2004)]{werner:04} that allowed the
detection of magnesium, with an abundance of about 2\%
(Fig.\,\ref{fig:h1504}). Another strong argument in favor of this idea would be
the detection of sodium, which would be direct evidence for C-burning. Stellar
models predict that the $^{23}$Na abundance at the bottom of the C/O envelope is
comparable to that of neon (main isotope $^{20}$Ne) and magnesium
($^{24,25,26}$Mg, \cite[Iben \etal 1997]{iben:97}). We are analyzing new UV
spectra taken with HST/COS aiming at the abundance determinations for Mg and Na
(Werner \etal 2010).

It could turn out that \hh\ is one of the ``heavy-weight'' intermediate-mass
stars (8\,M$_{\odot}\ \lappr$ $ M\ \lappr$\ 10\,M$_{\odot}$) that form white
dwarfs with electron-degenerate O--Ne--Mg cores.  At present it is uncertain
under which circumstances super-AGB stars (i.e. the massive counterparts of AGB
stars that ignite carbon but do not proceed to further stages of nuclear
burning) produce O--Ne--Mg WDs or explode as electron-capture SNe producing NSs
(e.g. \cite[Siess 2007]{si07}). This uncertainty mainly arises from modeling
uncertainties in mass-loss and mixing processes.  The possibility that \hh\ is a
O--Ne--Mg WD is remarkable, because evidence for the existence of such objects
is rather scarce \cite[(Weidemann 2003)]{weidemann:03}. Evidence from single
massive WDs is weak, and the most convincing cases are WDs in binary
systems. Strong Ne overabundances are found in novae \cite[(Livio \& Truran
1994)]{livio:94} or in eroded WD cores in LMXBs \cite[(Juett \etal
2001)]{juett:01}. If a high Na abundance is found in \hh, then this would be the
most compelling case for the existence of a single O--Ne--Mg WD, i.e. a post
super-AGB star.  \hh\ would then also challenge stellar evolution theory
relevant for super-AGB stars, because it cannot explain how the star has lost
its H-rich and He-rich envelopes and why it exposes its metallic core.

At any rate, one may wonder where potential progenitors and successors with
H1504-like chemistry are. All [WR] and PG1159 stars analyzed so far are still
having significant amounts of helium in their atmospheres (30-50\%) and always C
$\gg$ O. Recently, a group of relatively cool white dwarfs (\Teff\ $\approx$
20\,000\,K) with almost pure carbon atmospheres were discovered (so-called ``hot
DQs''; \cite[Dufour \etal 2007, 2010]{dufour:07,dufour:11}). They may be evolved
H1504-like stars because one can expect that at some point in future evolution
carbon as lightest element will float on top of \hh.

Recently, \cite[G\"ansicke \etal (2010)]{gae10} discovered a new class of
two O-rich white dwarfs (He-dominated atmospheres with O/He\,$\approx$\,0.01 and
O\,$>$\,C, by number, and \Teff\ around 10\,000\,K) that could be O--Ne--Mg
white dwarfs.

\section{Summary and conclusion}

The rich diversity of hydrogen-deficient central stars of planetary nebulae and
related objects is not well understood. The majority consists of [WC] and PG1159
stars which are thought to form an evolutionary sequence that is caused by a
late helium-shell flash in post-AGB stars. The helium-dominated O(He) stars
suggest that there exists at least one additional H-deficient sequence that is
caused by a double-degenerate merger. It is possible that the O(He) stars are
the progeny of R CrB stars.

There is growing evidence that there exists a H-deficient post-super AGB
white-dwarf cooling sequence consisting of remnants with O--Ne--Mg cores. \hh\
with its C-O dominated atmosphere might mark the hot end of this sequence while
the recently discovered groups of C-dominated ``hot DQs'' and He-dominated
O-rich white dwarfs occupy cooler regions. The origin of H-deficiency in
post-super AGB stars is unknown. There is no obvious hint as to which central
stars of planetary nebulae could be related to these chemically peculiar white
dwarfs.  \newpage
\noindent {\bf Acknowledgements} \quad I thank Thomas Rauch and Jeff Kruk for
their enduring collaboration. HST data analysis in T\"ubingen is supported by
the German Ministry of Education and Research through the German Aerospace
Center (grant 05\,OR\,0806).

\end{document}